\documentclass[fleqn,usenatbib]{mnras}

\usepackage{newtxtext,newtxmath}

\usepackage[T1]{fontenc}

\DeclareRobustCommand{\VAN}[3]{#2}
\let\VANthebibliography\thebibliography
\def\thebibliography{\DeclareRobustCommand{\VAN}[3]{##3}\VANthebibliography}

\usepackage{graphicx}	
\usepackage{amsmath}	

\title[Constraining the Lifespan of Intelligent Technological Civilization in the Galaxy]{\textbf{Constraints on} the Lifespan of Intelligent Technological \textbf{Civilizations} in the Galaxy}

\author[Sohrab Rahvar and Shahin Rouhani]{
Sohrab Rahvar,$^{1,2}$\thanks{E-mail: rahvar@sharif.edu}
Shahin Rouhani,$^{1,2}$\thanks{E-mail: srouhani@sharif.edu}
\\
$^{1}$Department of Physics, Sharif University of Technology, Azadi ave. Tehran 11365-9161, Iran\\
$^{2}$Research Center for High Energy Physics, Sharif University of Technology, Tehran, Iran\\
}

\date{Accepted XXX. Received YYY; in original form ZZZ}

\pubyear{\the\year{}}

\begin{document}
\label{firstpage}
\pagerange{\pageref{firstpage}--\pageref{lastpage}}
\maketitle

\begin{abstract}
In this work, we explore constraints on the emergence and longevity of technologically intelligent civilizations in our Galaxy, considering the Fermi paradox. We argue that under optimistic assumptions about the probability of life and intelligence emerging on Earth-like planets, the absence of contact with extraterrestrial civilizations imposes limits on their lifespan. Our analysis suggests that if intelligent life is common, technological civilizations must be relatively short-lived, with lifetimes constrained to $\lesssim 5\times10^3$ years under our most optimistic scenario. Considering electromagnetic communication, we note that our current light cone encompasses the entire Galactic history over the past $\sim 10^5$ years, making the lack of detected signals particularly puzzling for long-lived civilizations. We emphasize that these results should be interpreted as upper bounds derived from the Fermi paradox, not as predictions of actual lifespans.
\end{abstract}




\section{Introduction}
\label{sec:introduction}

Enrico Fermi posed a compelling question: If intelligent life and advanced civilizations exist within our galaxy, why have we not encountered any evidence of them? Considering a hypothetical advanced civilization that existed in our galaxy hundreds of millions of years ago, a mere fraction of a star's lifespan. It would have explored the galaxy within that time, and we might have already interacted with them, or they could still be present on Earth. Alternatively, they would have constructed the Dyson sphere around their Sun as manifestation of their advanced civilization, and we should have received signals of such a sphere \citep{dyson}. What chance do they stand of finding us? We assume that the advanced civilization has spaceships that can travel at one-tenth of the speed of light.  Then, a simple calculation for the upper limit is based on the diffusion process in which a single spaceship after $N$ steps can reach the distance of $\ell =\sqrt{N}l_s$. Taking the size of Galaxy $\ell \simeq 10^5$ {\it lyr} and the distance between the stars $l_s\simeq1$ {\it lyr}. Then after $N=10^{10}$ steps, this spaceship can explore $10\%$ of the stars in the Galaxy, assuming that Galaxy has $10^{11}$ stars, then traveling at the speed of $0.1c$, it takes $10^{11}$ years. So, the exploration time should be less than $t_{exp}<10^{11}~\text{yr}$, if they use one spaceship. However, using a large number of spaceships can reduce exploration time by orders of magnitude.  For instance, if we have a large number of spaceships, then the minimum time scale would be just time need to traverse the radius of the Galaxy at the speed of $0.1c$, resulting in $10^6$yr. Then we can conclude that the exploration time would be in the range of 
$10^6~\text{yr}<t_{exp}<10^{11}~\text{yr}$. 

One must also consider electromagnetic communication rather than physical travel. The propagation time for signals across the Galaxy is $\sim 10^5$ years, but this must be considered in the context of our light cone. Since we observe the Galaxy through our past light cone, we effectively have access to the entire Galactic history over the past $\sim 10^5$ years. This means that any technological civilization that broadcasted electromagnetic signals anywhere in the Galaxy within the last $10^5$ years should, in principle, have those signals reaching us now. The absence of detected artificial signals despite decades of SETI searches and all-sky surveys therefore imposes particularly strong constraints on the prevalence and longevity of technological civilizations. For a civilization with lifespan $L$, if $L \gtrsim 10^5$ years and it broadcasts detectable signals for a significant fraction of its existence, we would expect to see evidence of its electromagnetic activity within our past light cone. This electromagnetic argument complements the physical travel constraints and, given our current observational capabilities, may provide even tighter bounds on civilization lifetimes.

One possible explanation for Fermi's enigma is that we may be the only and first advanced civilization currently existing in our Galaxy or that we are the most advanced civilization.
Some authors have taken this line of argument, claiming that the chances of an intelligent life form appearing anywhere is extremely small. So our existence is very rare in the galaxy \citep{rareearth}. Alternatively, the advanced civilizations may have happened several times in our Galaxy but the lifespan of a civilization on a planet might be too short to achieve either the advanced manifestation of the civilization, the Dyson sphere or to explore the galaxy. It's also conceivable that other extra-terrestrial civilizations self-destructed through wars, overuse of resources, tampering with the climatic system, or simply destroyed by a viral pandemic, leading to their planet's destruction and the end of their existence. The possible causes of human extinction have been listed as Asteroid Impact, Super volcano Eruptions, Gamma-Ray Bursts, Climate Change, Pandemics, Nuclear War, Technological origin (artificial intelligence, rogue biotechnology), Cosmic Events and societal collapse \citep{2019BBCFutureCollapse,martin_rees}. It is conceivable that any technologically advanced civilization should face the same threats.

Many civilizations have suffered such a fate, regardless of their size or complexity. Most never recovered, such as the Western and Eastern Roman Empires, the Maya, and the civilization of Easter Island \citep{2019BBCFutureCollapse,BROZOVIC2023103075}. We note that the collapse of Easter Island civilization is attributed by some sources primarily to external factors including contact with Europeans, though resource depletion may have played a contributing role. However, some of them later revived and transformed, such as China, Greece, Egypt, and Iran.  All these happened when the world was less connected; in today's highly connected world it is possible that the entire planet could have such a fate.

Let us start with two questions on it: Once a habitable planet is formed, how likely is it for life to form? 
 Secondly once life is formed on a planet how long does it take to arrive at an intelligent/communicating species capable of a technologically advanced civilization?

Let $N_{EH}$ be the number of earth-like habitable planets in our galaxy. Having $N_\star$ as the total number of stars in Galactic disk, $\eta_\star$ fraction of the stars that sustain habitable planet and $\eta_{EH}$ is the fraction of Earth-like habitable planets around such stars. The number of habitable Earth-like planets in our Galaxy would be: 
\begin{equation}N_{EH} = N_\star\times\eta_\star\times\eta_{EH}.
\end{equation}

Various estimations suggest that certain types of stars (specifically types O, B, and A) can not support life \citep{supportlife}. The large stars suffer from the short-life time. Also some authors argue that dwarf stars have the problem of the tidal locking of planets in the habitable zone \citep{Mdwarf,2024arXiv241205002S}. However recent observations of K2-18b by the James-Webb telescope, a super-earth mass planet orbiting about a red-dwarf, exhibit the possible signatures of life \citep{2025ApJ...983L..40M}. Some authors consider an extended orbit for the tidally locked planets where parts of planets still satisfy the habitability condition \citep{2016ApJ...819...84K}. So we take M, K, G and F type stars in our analysis.

Also, a fraction of binary star systems can support sustainable conditions on the planets, as the chaotic orbits of planets around binary stars can produce a strong and long-term change in the climate of the planet \citep{10.1093/mnras/sts257}. We also take only the third generation of stars, as the first and second generations lack sufficient metals to form telluric (rocky) planets. We take stars in the disk in contrast to the dense area of the Galactic bulge, which has a destructive perturbation effect on the orbit of planets around the stars \citep{doi:10.1142/S0218271821500632}. Additionally, we must consider the migration of Jupiter-like planets towards their parent stars, which can impact the rocky planets within the snow line. Considering all factors, we adopt recent analysis on the Earth-like habitable planets in our Galaxy, $N_{EH} =2.5^{+71.6}_{-2.4}\times 10^5$ \citep{2024arXiv241205002S}. Note that the upper error bar here is not a measurement but reflects the uncertainty range; we adopt $N_{EH} \sim 10^5-10^7$ as a plausible interval. Other studies also estimate a larger value of $N_{EH} = 4\times 10^6$ \citep{2006ilu..book.....U}.

Given that the galaxy is polluted with life molecules \citep{2022FrASS...987567G}, originating from chemical reactions in the interstellar medium or electric discharges in the atmosphere of planets \citep{Urey} for a habitable planet, we expect to have life with probability of one. To be clear, we expect some form of molecule capable of repeating itself to land on the planet, then the workings of evolution will lead to the early starters of life i.e., cells. This process, known as abiogenesis (the origin of life from non-living matter), is the most plausible precursor of life, but it does need water. Since the formation of earth crust cooled down, the first cellular life happened 3.8 Gyr ago. We conclude that life forms rapidly on habitable planets.    

The next question then becomes how long it will take for an intelligent species to evolve, and then how long it takes for an intelligent communicating ( IC ) species to evolve. 
This is a complicated question. If we accept that evolution tends to increase complexity \citep{McShea, McShea2021,McShea2010,McShea2003} , the appearance of intelligent beings and then intelligence plus written record, is a certainty and just a matter of time. That the natural flow of evolution is toward greater complexity has been argued by many authors; for instance, 
\cite{Bottcher2016} proposes a definition of molecular complexity and a definition of DNA complexity coding for that molecule. This works well; the two complexities are linearly related before the appearance of eukaryotes. Evidence for an increase in complexity can be observed across many genera.  The change from prokaryotes to eukaryotes is itself understood as an increase in complexity \citep{2025PNAS..122...13M}
which took almost 2 Gyr to happen. If life on earth is a guide we can even derive a rate of change of complexity \citep{Bottcher2016}.

Life appeared on Earth around 4 billion years ago. But those first organisms were very simple cells \citep{2022absc.conf21705E}; it would be another 1.4 billion years before multicellular life appeared. Animals probably evolved even more recently, around 800 million years ago \citep{Brain_Prave_Hoffmann_Fallick_Botha_Herd_Sturrock_Young_Condon_Allison_2012}.
At this rate it takes 2 Gyr to reach the algorithmic revolution needed for eukaryotes and then 1.5 Gyr for the appearance of homo sapiens. The other revolutions which happened are as Cambrian star speciation ($\sim$ 500 Myr ago), formation of oxygen rich atmosphere ($\sim$ 300 Myr ago), the Permian mass extinction around 250 Myr ago (most likely caused by an asteroid impact) and  Jurassic 65Myr by an asteroid impact, appearance of mammals, $\sim$ 200 Myr. Other events which seem to be associated with evolutionary jumps are the apes diverged from monkeys around 25 Myr ago, evolution of bipedalism around 6 Myr. Increase in brain size and tool use began around 2.6 Myr ago by homo habilis, control of fire 800 Kyr and Homo sapiens emerged around 300 Kyr ago.

Start by assuming that a rocky planet is in a habitable zone (and has water). We ask how long it will be before life is seeded on it. Judging by the example of life on Earth, with simple cells existing $\sim 4$ Gyr ago, abiogenesis must have occurred even earlier. This suggests that once a planet becomes habitable, life of interstellar or internal origin can begin quickly. Then, several steps are required before intelligent life appears, such as the development of eukaryotic cells, multicellularity, sexual reproduction, etc. The timescales for these transitions have been estimated, and even with many steps, the total time is finite \citep{2012PNAS..109..395S}. Considering that over $99\%$ of all species that have ever lived on Earth are estimated to be extinct \citep{Stearns1999}, and given that there are approximately 11 million extant species \citep{Larsen2017}, the total number of species that have ever existed is on the order of 1 billion. This estimate aligns with statistical models of biodiversity over geological timescales that account for speciation and extinction rates. Then, note that Homo sapiens were not the only intelligent species; even within the genus Homo, there were several coexisting species such as Denisovans and Neanderthals \citep{2024Natur.632..108X}.

The next jumps are not just genetic; they include the Neolithic Revolution 12,000 years ago, written records, and then industrialization, the scientific revolution, etc. These may have happened randomly or as some kind of feedback method, which, by the way, has not happened for other intelligent species on Earth. We are then at a difficult point and can only assume that complexity increases with time. Evolution will eventually lead to an intelligent industrial being, with a timescale of 4 Gyrs since the original inception. 

 Assuming that the earth is typical, one may say that the average time it takes for IC to appear at the end of some stochastic search path in the configuration space of all species is $4$ billion years. This timescale appears arbitrary since what matters in evolution is the generation time. And each exoplanet may have its own time scale based on its specification and parent star. However the generation time may not only depends on the length of the year (one loop around the star) but also on the rotation speed around the axis of the planet (length of day and night cycle) and many other parameters such as average temperature. Thus in summary we cannot say that 4Gyr is not the typical evolution time for an IC civilization to appear. 

 Here we assume that any habitable planet will be seeded by life either internally or by an external agent, assuming that increasing complexity will eventually generate an industrialized intelligent life. Let us consider the probability of this event happening as $f$. In the following we give a more precise definition of $f$ and we will estimate constraints on the lifetime that this intelligent life can survive on a planet.

Let $f_L$ represent the fraction of stars with habitable planets that harbor life,  $f_I$ the fraction with intelligent life, and $f_T$  the fraction with technological life. If we denote $L$ as the lifespan of technological life and $L_s$ as the life-time of a star, the Drake equation gives us the number of intelligent civilizations in our galaxy as \citep{drake}
\begin{equation}
N=N_{EH} \times f_L \times f_I \times f_T\times L/L_s,
\end{equation}
where we can simplify this equation by defining $f=f_L \times f_I \times f_T$ leading to: $N=N_{EH} \times f\times L/L_s$ where we assume that intelligent life only once appears in the history of a planet.

On Earth, technological societies have existed for only about $200$ yrs, which we can hypothesize that they might endure for $1,000$ or $10,000$ years more, despite challenges. Let’s set the expected life span of such a society as $L=10^n$ years. For the case of the sun, earth's life will last at most up to $1.3$Gyr \citep{2015JGRD..120.5775W}. This is due to a change in the flux of the sun as a G-type star. Assuming that life emerged $3.8$ Gyr ago, then the total lifetime for the sun that supports life on Earth would be $L_s \simeq 5$ Gyr. Here, we ignore the possibility that an advance civilization would be able to alter the orbit of its planet for extending the habitability of a planet \citep{rahvar}. 

\begin{figure}
\includegraphics[width=\columnwidth]{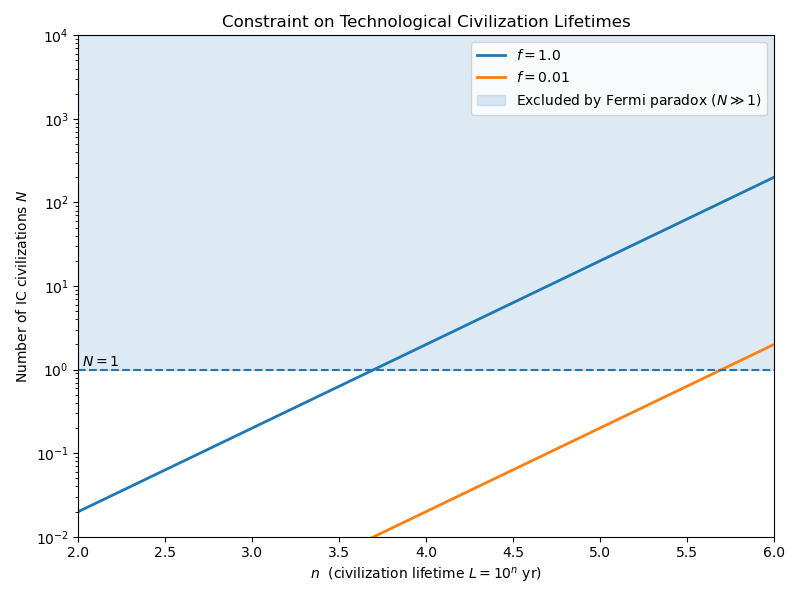}
    \caption{Constraint on the lifetime $L=10^n$ of intelligent technological civilizations derived from the Fermi paradox.
The expected number of coexisting civilizations $N$ is shown as a function of the civilization lifetime for different assumptions about the probability $f$ that a habitable planet produces a technological civilization.
The shaded region corresponds to values for which $N \gg 1$, which are observationally excluded by the absence of detected extraterrestrial civilizations.
The dashed line indicates the boundary $N=1$ and is shown for reference.}
    \label{fig_1}
\end{figure}

Taking $N_{EH} \sim 10^6$ in our Galaxy and 
substituting these values, we find: $N\simeq f/5\times 10^{(n-3)}$.  Figure (\ref{fig_1}) represents $N$ as a function of $n$ for two values of $f$.   The value of $f$ is uncertain. Assuming an optimal value of $f=1$, implies that intelligent life with technology will emerge on all Earth-like habitable planets. This results in a relation between $N$ and $n$, $N\simeq 1/5\times 10^{(n-3)}$. If we set $n=3.7$  we end up with N=1 that is, we are alone in the galaxy at the present time. Or in other words, there should be at least one IC civilization in our Galaxy at any given time. This is a rough estimation from the Poisson process. Under these optimistic assumptions, the absence of contact with other civilizations implies a statistical upper bound on the typical lifespan of technological civilizations of $\sim 5000$ years, corresponding to the requirement that the expected number of coexisting civilizations satisfies $N\lesssim1$. We emphasize that this is not a prediction but a constraint derived from the Fermi paradox; actual lifespans could be much shorter. For instance, human civilization currently has the means to destroy itself within centuries, far below this bound. Conversely, a smaller value of $f$ would allow for a longer lifespan for technological civilizations still resolving the Fermi paradox. Thus, we can establish a constraint for the lifespan of advanced technological civilization as:  $f<1$ solves the Fermi paradox with the condition of $L>5000$ yr. In our analysis, we assumed that all Earth-like habitable planets would end up having intelligent life. Now we want to investigate the likelihood of this event.

Taking into account a lifetime of $5\times 10^3$ yr for an IC civilization, we pose the question of up to what distance this civilization can diffuse in our Galaxy.  Assuming Voyager probes as an example, it took $\sim 50$ yr to reach the distance of $\sim 100$ a.u. In $5000$ yrs, assuming linear displacement, that will reach to the distance of $0.1$ pc, which is not enough to reach a nearby star. However this is a lower limit and an advanced civilization may manufacture a spaceship that travels at a fraction of the speed of light, let us take $0.1$c. Then, with approximately $N_{EH} =10^6$ stars having Earth-like planet, the typical distance between two Earth-like planets would be in the order of $10^2$ pc, meaning that our assumed spaceship will take $3000$ yrs, which in the order of $5000$ yrs. However there is sociological and psychological challenges for such a long space travel.

Let us extend this analysis for the case of $N=2$, which deals with the Fermi paradox. Why has another advance civilization  has not been discovered us yet. This is true at least during the documented history of human.  
Using this null event, we may put constrain on the age of an advanced civilization. In the Drake's equation, $N = N_{EH} f L/Ls$, we assume the maximum value of $f=1$ 
as the outcome of complexity in an Earth-like planet. Let us model our Galaxy is an almost 2-dimensional structure with a radius of D, then the column number density of intelligent life would be $\Sigma = N/D^2$. So the typical distance between any Intelligent life would be $d = D/\sqrt{N}$. Assuming advanced civilization can travel at ten percent of speed of light, the time of finding another civilization inside the Galaxy would be $t_f = d/0.1c$. Then, using the radius of our Galaxy in the order of $D = 3\times 10^4$ ly, the time of finding is 
$$ t_f = 3\times 10^5/\sqrt{N} ~\text{yr}.$$
In order to have at least two advanced civilization at one time we let $N=2$, which results in $t_f \sim 200~ \text{kyr}$. So to have one visit possible, the lifetime of advanced intelligent life should be at least $200$~kyr. We can imagine that an advanced civilization have already visited us as the homosapien, however our historical memory dates back at least to $10$ kyr. Having no visit or artifacts of possible genetic engineering by visitors implies that lifetime of an advanced civilization should be smaller than $t_f$. 

Considering electromagnetic communication, our light cone provides a complete census of Galactic activity over the past $\sim 10^5$ years. The absence of detected artificial signals despite decades of SETI searches strengthens the argument that technological civilizations must either be rare, have very short broadcasting phases, or choose not to communicate. This electromagnetic constraint is particularly powerful because it doesn't require civilizations to send probes; it only requires that they broadcast detectable signals at some point during their existence.

It is important to emphasize that the two characteristic timescales derived above correspond to distinct logical regimes. 
The $\sim5\times10^3$~yr bound arises from requiring that the expected number of coexisting technological civilizations be $N\lesssim1$, under optimistic assumptions about habitability and the emergence of intelligence.
In contrast, the $\sim10^5$~yr timescale characterizes the minimum lifetime required for detection or contact \emph{given that} $N\geq2$ civilizations coexist in the Galaxy.
Technological civilizations with lifetimes in the intermediate range $5\times10^3~\text{yr}\lesssim L\lesssim10^5~\text{yr}$ are therefore fully consistent with the Fermi paradox: multiple civilizations may exist sequentially or even overlap in time, yet still fail to detect one another due to insufficient longevity for communication or travel.
Only for lifetimes significantly exceeding $\sim10^5$~yr does the absence of evidence become increasingly difficult to reconcile under optimistic assumptions.

Summarizing, the absence of evidence for any other IC implies many conclusions. (i) We have not yet observed it. 
(ii) there exist one IC in the Galaxy at any one time (i.e. $N=1$ this being us), this implies that such a being should live  $L\lesssim 5000$ yr, if we adopt the optimistic case $f\simeq 1$. 
(iii) There exist two IC and probability of discovering each other either by the electromagnetic signals or direct contact needs a life-time having more than $L\sim 10^5$yr.  

We conclude that an IC civilization may die out or collapse to a pre-technological state before it manages to diffuse into the Galaxy.
This might be a resolution for the Fermi's paradox. However, we stress that these conclusions are derived under specific assumptions and represent constraints rather than predictions. Other resolutions to the Fermi paradox (e.g., civilizations choose not to communicate, or we are early) remain viable alternatives.

\section*{DATA AVAILABILITY}
No observational and simulation data is generated in this article. 



\bibliographystyle{mnras}
\bibliography{example} 
\end{document}